\documentclass[pra,aps,twocolumn,amsmath,amssymb,showkeys]{revtex4}

\newcommand{\hh}{\mathcal{H}}
\newcommand{\lnp}{\mathcal{L}}
\newcommand{\lsa}{\mathcal{L}_{s.a.}}
\newcommand{\lsp}{\mathcal{L}_{+}}
\newcommand{\bro}{\boldsymbol{\rho}}
\newcommand{\bsg}{\boldsymbol{\sigma}}
\newcommand{\blm}{\boldsymbol{\lambda}}

\newcommand{\pen}{\openone}

\newcommand{\ax}{\mathsf{X}}
\newcommand{\ay}{\mathsf{Y}}
\newcommand{\mm}{\mathsf{M}}
\newcommand{\nm}{\mathsf{N}}
\newcommand{\wnm}{\widetilde{\mathsf{N}}}
\newcommand{\tr}{\mathrm{tr}}
\newcommand{\mc}{\mathcal{M}}
\newcommand{\nc}{\mathcal{N}}

\newcommand{\wta}{\widetilde{a}}
\newcommand{\wtb}{\widetilde{b}}
\newcommand{\wrn}{\widetilde{R}}
\newcommand{\wrh}{\widetilde{H}}

\begin{document}
\clearpage
\preprint{}

\title{Notes on general SIC-POVMs}

\author{Alexey E. Rastegin}
\affiliation{Department of Theoretical Physics, Irkutsk State University,
Gagarin Bv. 20, Irkutsk 664003, Russia, e-mail: {\tt alexrastegin@mail.ru}}

\begin{abstract}
An unavoidable task in quantum information processing is how to
obtain data about the state of an individual system by suitable
measurements. Informationally complete
measurements are relevant in quantum state tomography, quantum
cryptography, quantum cloning, and other questions. Symmetric
informationally complete measurements (SIC-POVMs) form an
especially important class of such measurements. We formulate some
novel properties and relations for general SIC-POVMs in a
finite-dimensional Hilbert space. For a given density matrix and
any general SIC-POVM, the so-called index of coincidence of
generated probability distribution is exactly calculated. Using
this result, we obtain state-dependent entropic bounds for a
single general SIC-POVM. Lower entropic bounds are derived in
terms of the R\'{e}nyi $\alpha$-entropies for
$\alpha\in[2;\infty)$ and the Tsallis $\alpha$-entropies for
$\alpha\in(0;2]$. A lower bound on the min-entropy of a SIC-POVM
is separately examined. For a pair of general SIC-POVMs, entropic
uncertainty relations of the Maassen--Uffink type are considered.
\end{abstract}


\maketitle

\pagenumbering{arabic}
\setcounter{page}{1}

\section{Introduction}\label{sec1}

The quantum information science uses quantum states and effects as
tools for information processing \cite{nielsen}. At the final
stage of any protocol, some measurements are required. Hence, we
ask how to obtain information about the state of a quantum system.
Informationally complete measurements have found to be useful in
many issues. Among numerous methods for retrieving information on
the state, the informationally complete measurements
\cite{prug77,busch91} seem to be the most versatile. Especially
interesting cases are when the measurement is symmetric
\cite{rbsc04} or covariant with respect to a group of physical
transformations \cite{gdps04,dps04}. Symmetric informationally
complete (SIC) measurements are the subject of active research.
Despite of simple definition, SIC-POVMs are difficult to
construct. Studies of symmetric POVMs are connected with many
mathematical problems \cite{afz13,anw14}. They are intimately
related to problem of building a complete set of mutually unbiased
bases (MUBs) \cite{bz10}. Weyl--Heisenberg (WH) covariant SIC-sets
of states were examined in \cite{adf07}. Tight informationally
complete measurements were introduced in \cite{scott06}.

In their original version, SIC-POVMs are assumed to be constructed
of only rank-one elements. Some concrete examples in low
dimensions are discussed in  \cite{rbsc04}. For quantum
tomography, rank-one SIC-POVMs are maximally efficient at
estimating the quantum state \cite{rbsc04}. The seamy side is that
such a measurement erases the original state of the system being
measured. These reasons pertain to situation, when some unknown
state is the subject of tomography and also post-tomography
processing. It is typical in such a case that only a part of the
total system is measured through the tomography process. As a
rule, there is a trade-off between efficiency of used measurement
and disturbance, which will influence on further stages. In this
regard, other versions of SIC-POVMs are of interest. An
approximate version of rank-one SIC-POVMs were examined in
\cite{krsw05}. General SIC-POVMs with elements of any rank are
considered in \cite{appl07,kalv13}. It has recently been shown
that general SIC-POVMs exist in all dimensions \cite{ggour13}.
Explicit constructions for such POVMs and their dual bases have
been presented in \cite{ggour13}.

The aim of the present work is to study some generic properties of
general symmetric informationally complete measurements. The
preliminary material is reviewed in Section \ref{sec2}. In
particular, dual bases of informationally complete POVMs are
briefly considered. In Section \ref{sec3}, we exactly calculate
the so-called index of coincidence of probability distribution
obtained with a general SIC-POVM and arbitrary state. Uncertainty
bounds for a single general SIC-POVM are considered in Section
\ref{sec4}. For these purposes, we respectively use the Tsallis
and R\'{e}nyi entropies, including the so-called min-entropy.
Within the Tsallis formulation, we obtain uncertainty relations in
the case of detection inefficiencies. In Section \ref{sec6},
uncertainty relations of the Maassen--Uffink type are obtained for
a pair of general SIC-POVMs. In Section \ref{sec7}, we conclude
the paper with a summary of results.

\section{Notation}\label{sec2}

Let $\lnp(\hh)$ be the space of linear operators on
$d$-dimensional Hilbert space $\hh$. By $\lsa(\hh)$ and
$\lsp(\hh)$, we respectively mean the space of Hermitian operators
on $\hh$ and the set of positive ones. For two operators
$\ax,\ay\in{\mathcal{L}}(\hh)$, their Hilbert--Schmidt inner
product is defined by \cite{watrous1}
\begin{equation}
\langle\ax{\,},\ay\rangle_{\mathrm{hs}}:=\tr(\ax^{\dagger}\ay)
\ . \label{hsdef}
\end{equation}
The inner product (\ref{hsdef}) induces the Frobenius norm, also
called the Hilbert--Schmidt norm:
\begin{equation}
\|\ax\|_{2}:=\langle\ax{\,},\ax\rangle_{\mathrm{hs}}^{1/2}
=\tr\bigl(\ax^{\dagger}\ax\bigr)^{1/2}
\ . \label{hsmnm}
\end{equation}
To each $\ax\in\lnp(\hh)$, we assign positive operator
$|\ax|:=\sqrt{\ax^{\dagger}\ax}$. The singular values
$\sigma_{i}(\ax)$ of arbitrary $\ax\in\lnp(\hh)$ are defined as
the eigenvalues of $|\ax|\in\lsp(\hh)$. Then the Schatten $q$-norm
is introduced for all $q\in[1;\infty]$ as
\begin{equation}
\|\ax\|_{q}:=\left(\sum\nolimits_{i=1}^{d}\sigma_{i}(\ax)^q\right)^{1/q}
{\,}. \label{shndf}
\end{equation}
The family (\ref{shndf}) gives the trace norm
$\|\ax\|_{1}=\tr|\ax|$ for $q=1$, the Frobenius norm (\ref{hsmnm})
for $q=2$, and the spectral norm
$\|\ax\|_{\infty}=\max\bigl\{\sigma_{i}(\ax):{\>}1\leq{i}\leq{d}\bigr\}$
for $q=\infty$. These norms and relations between them have found
to be useful in various questions of quantum information
\cite{watrous2,rast103,rcun12}. For each $q\in[1;\infty]$ and
$\ax,\ay,\in{\cal{L}}(\hh)$, we have \cite{watrous1}
\begin{equation}
\|\ax\ay\|_{q}\leq\|\ax\|_{\infty}{\,}\|\ay\|_{q}
\ . \label{schqq}
\end{equation}
We will used (\ref{schqq}) in section \ref{sec6}. For all
$q>p\geq1$ and arbitrary $\ax\in\lnp(\hh)$, we also have
\begin{equation}
\|\ax\|_{q}\leq\|\ax\|_{p}
\ . \label{npqr}
\end{equation}
This relation is actually a consequence of theorem 19 of the
classical book \cite{hardy}.

A density matrix $\bro\in\lsp(\hh)$ has unit trace, i.e.
$\tr(\bro)=1$. Generalized quantum measurements are commonly
described within the POVM formalism \cite{peresq}. Let
$\nc=\{\nm_{j}\}$ be a set of elements $\nm_{j}\in\lsp(\hh)$,
satisfying the completeness relation
\begin{equation}
\sum\nolimits_{j} \nm_{j}=\pen
\ . \label{cmprl}
\end{equation}
Here, the $\pen$ denotes the identity operator on $\hh$. This set
$\nc=\{\nm_{j}\}$ is a positive operator-valued measure (POVM).
Consider some POVM with $d^{2}$ elements $\nm_{j}$, which satisfy
the following two conditions. First, for all $j=1,\ldots,d^{2}$ we
have
\begin{equation}
\langle\nm_{j}{\,},\nm_{j}\rangle_{\mathrm{hs}}=a
\ . \label{ficnd}
\end{equation}
This condition can be rewritten as $\|\nm_{j}\|_{2}=\sqrt{a}$.
Second, the pairwise inner products are all symmetrical, namely
\begin{equation}
\langle\nm_{j}{\,},\nm_{k}\rangle_{\mathrm{hs}}=b
\qquad (j\neq{k})
\ . \label{secnd}
\end{equation}
Then the POVM $\nc=\{\nm_{j}\}$ is a general SIC-POVM.
Combining
$\langle\pen{\,},\pen\rangle_{\rm{hs}}=d$ with (\ref{ficnd})
and (\ref{secnd}) finally gives the relation \cite{ggour13}
\begin{equation}
b=\frac{1-ad}{d(d^{2}-1)}
\ . \label{bvia}
\end{equation}
Further, we obtain $\tr(\nm_{j})=d^{-1}$ for all
$j=1,\ldots,d^{2}$. Therefore, the value $a$ is the only parameter
characterizing the type of a general SIC-POVM. This parameter is
restricted as \cite{ggour13}
\begin{equation}
d^{-3}<{a}\leq{d}^{-2}
\ . \label{resa}
\end{equation}
The lower bound $a=d^{-3}$ is reached in the case
$\nm_{j}=d^{-2}\pen$, which does not give an informationally
complete POVM. The upper bound $a=d^{-2}$ is achieved, if and only
if the POVM elements are all rank-one. The latter is actually the
case of usual SIC-POVMs, when each element is represented in terms
of the corresponding unit vector as
\begin{equation}
\nm_{j}=d^{-1}|\phi_{j}\rangle\langle\phi_{j}|
\ . \label{raon}
\end{equation}
The formulas (\ref{secnd}) and (\ref{bvia}) then result in
\begin{equation}
\bigl|\langle\phi_{j}|\phi_{k}\rangle\bigr|=\frac{1}{\sqrt{d+1}}
\ . \label{raon1}
\end{equation}
In the further discussion, we will usually avoid both the least
values of the relation (\ref{resa}). That is, we exclude the case
$\nm_{j}=d^{-2}\pen$ as well as rank-one SIC-POVMs.

In the next sections, we will use the following properties of
general SIC-POVMs. First, the elements $\nm_{j}$ of a general
SIC-POVM form a basis in the space $\lsa(\hh)$ \cite{ggour13}.
Second, each operator $\ax\in\lsa(\hh)$ can be represented in
terms of elements of the dual basis as
\begin{equation}
\ax=\sum\nolimits_{j=1}^{d^{2}}
\langle\nm_{j}{\,},\ax\rangle_{\mathrm{hs}}{\,}\wnm_{j}
\ . \label{axdb}
\end{equation}
Here, the dual basis $\bigl\{\wnm_{j}\bigr\}$ is a basis in
$\lsa(\hh)$ such that
\begin{equation}
\langle\nm_{j}{\,},\wnm_{k}\rangle_{\mathrm{hs}}=\delta_{jk}
\qquad \forall{\>}j,k\in\{1,\ldots,d^{2}\}
\ . \label{dbdf}
\end{equation}
When some basis in $\lsa(\hh)$ is formed by positive operators,
its dual basis cannot consist of only positive elements
\cite{spek08,fere08}. For a general SIC-POVM $\{\nm_{j}\}$, the dual basis is comprised by operators
\cite{ggour13}
\begin{equation}
\wnm_{j}=\frac{d}{ad^{3}-1}{\,}
\Bigl((d^{2}-1)\nm_{j}-(1-ad)\pen\Bigr)
\ . \label{dbsic}
\end{equation}
For a usual SIC-POVM with elements (\ref{raon}), the formula
(\ref{dbsic}) is reduced to
$\wnm_{j}=(d+1)|\phi_{j}\rangle\langle\phi_{j}|-\pen$. Explicit
constructions and related properties of general SIC-POVMs are
presented in \cite{ggour13}.

\section{Index of coincidence}\label{sec3}

In this section, we calculate the so-called index of coincidence
of probability distribution generated by a general SIC-POVM on any
mixed state. Using indices of coincidence, the writers of
\cite{molm09} derived entropic bounds for a set of several
mutually unbiased bases. For usual SIC-POVMs, this issue was
considered in \cite{rast13b}. The index of coincidence of
probability distribution $\{p_{j}\}$ is defined as
\cite{bengtsson,ht01}
\begin{equation}
C(p):=\sum\nolimits_{j} p_{j}^{2}
\ . \label{icdf}
\end{equation}
This quantity is often called purity \cite{bengtsson}, when
the probabilities are assumed to be eigenvalues of a density
matrix. Inverse of (\ref{icdf}) is known as the participation
number \cite{bengtsson}. If the pre-measurement state is described
by density matrix $\bro$, $j$-th outcome occurs with the
probability
\begin{equation}
p_{j}(\nc|\bro)=\tr(\nm_{j}\bro)
\ . \label{jopr}
\end{equation}
For the given SIC-POVM $\nc$ and state $\bro$, the quantity
$C(\nc|\bro)$ is obtained by substitution of probabilities
(\ref{jopr}) into (\ref{icdf}). The following statement takes
place.

\newtheorem{prpn1}{Proposition}
\begin{prpn1}\label{prp1}
Let general SIC-POVM $\nc$ be characterized by the parameter $a$
in the sense of (\ref{ficnd}). For arbitrary $\bro$, the index of
coincidence of generated probability distribution is equal to
\begin{equation}
C(\nc|\bro)=\frac{(ad^{3}-1){\,}\tr(\bro^{2})+d(1-ad)}{d(d^{2}-1)}
\ . \label{icpr}
\end{equation}
\end{prpn1}

{\bf Proof.} Using (\ref{axdb}) and (\ref{jopr}), we represent the
density matrix in the dual basis as
\begin{equation}
\bro=\sum\nolimits_{j=1}^{d^{2}}
p_{j}{\,}\wnm_{j}
\ . \label{rodb}
\end{equation}
For brevity, we put the quantities
$\wta=\langle\wnm_{j}{\,},\wnm_{j}\rangle_{\rm{hs}}$ and
$\wtb=\langle\wnm_{j}{\,},\wnm_{k}\rangle_{\rm{hs}}$ for
$j\neq{k}$. They are calculated by substitution of $\wnm_{j}$ and
$\wnm_{k}$ according to (\ref{dbsic}). The resulting expressions
are then written as
\begin{align}
&\wta-\wtb=\frac{d(d^{2}-1)}{ad^{3}-1}
\ . \label{abdf}\\
&\wtb=\frac{d(ad-1)}{ad^{3}-1}
\ , \label{wbdf}
\end{align}
Here, we used $\tr(\nm_{j})=d^{-1}$, the definitions (\ref{ficnd})
and (\ref{secnd}), and the condition (\ref{bvia}). Substituting
the right-hand side of (\ref{rodb}) into $\tr(\bro^{2})$ leads to
the formula
\begin{equation}
\tr(\bro^{2})=\wta{\,}C(p)+\wtb{\,}\sum\nolimits_{j\neq{k}}p_{j}p_{k}
=(\wta-\wtb){\,}{C}(p)+\wtb
\ , \label{rodb2}
\end{equation}
where the normalization condition was used at the last step.
Combining (\ref{rodb2}) with (\ref{abdf}) and (\ref{wbdf}) finally
leads to the claim (\ref{icpr}). $\square$

The statement of Proposition \ref{prp1} gives the expression for
the index of coincidence in terms of the parameter $a$ and
dimensionality $d$. For the completely mixed state
$\bro_{*}=\pen/d$ with $\tr(\bro_{*}^{2})=d^{-1}$, the formula
(\ref{icpr}) gives
\begin{equation}
C(\nc|\bro_{*})=d^{-2}
\ , \label{ccms}
\end{equation}
irrespectively to $a$. The right-hand side of (\ref{ccms}) is
valid, since $p_{j}(\nc|\bro_{*})=d^{-2}$ for any general
SIC-POVM. For a usual SIC-POVMs with only rank-one elements
(\ref{raon}), we substitute $a=d^{-2}$ into (\ref{icpr}) and
obtain
\begin{equation}
C(\nc|\bro)=\frac{\tr(\bro^{2})+1}{d(d+1)}
\ . \label{ucpr}
\end{equation}
The result (\ref{ucpr}) has been derived in \cite{rast13b} by
other method. For pure states, the numerator in the right-hand
side of (\ref{ucpr}) is equal to $2$. This pure-state case of
(\ref{ucpr}) was previously presented in \cite{rottler}. The
method of the paper  \cite{rottler} is based on the fact that the
unit vectors $|\phi_{j}\rangle$ form a spherical $2$-design. As
was noted in \cite{rast13b}, the result (\ref{ucpr}) is
significant from the viewpoint of applications in the entanglement
detection. It would be interesting to examine this question with
general SIC-POVMs. Indeed, general SIC-POVMs can be built for
arbitrary $d$.

In the case $d=2$, the right-hand side of (\ref{icpr}) can be
represented with use of the Bloch vector. Here, we denote the
identity $2\times2$-matrix by $\pen$ and the usual Pauli matrices
by $\bsg_{x}$, $\bsg_{y}$, and $\bsg_{z}$. Arbitrary density
matrix is written as
\begin{equation}
\bro=\frac{1}{2}{\>}\bigl(\pen+\vec{r}\cdot\vec{\bsg}\bigr)
{\>}. \label{bvr2}
\end{equation}
where $\vec{r}=(r_{x},r_{y},r_{z})$ is the Bloch vector.
Positivity of this matrix implies $r=|\vec{r}|\leq1$. Calculating
$\tr\bigl(\bro^{2}\bigr)=(1+r^{2})/2$, the index of coincidence is
equal to
\begin{equation}
C(\nc|\bro)=\sum_{j=1}^{4} p_{j}(\nc|\bro)^{2}=\frac{3+(8a-1)r^{2}}{12}
\ . \label{inc2}
\end{equation}
This result shows a dependence of $C(\nc|\bro)$ on the parameter
$a$ and the Bloch vector $\vec{r}$. For $a=1/4$, the formula
(\ref{inc2}) gives the fraction $(3+r^{2})/12$, which was already
noted in \cite{rast13b}. The Bloch-vector representation for
finite-level systems is one of important state representations
\cite{bengtsson}. Similarly to (\ref{inc2}), the formula
(\ref{icpr}) could be rewritten in terms of the generalized Bloch
vector of a $d$-level system. By $\blm_{n}\in\lsa(\hh)$, with
$n=1,\ldots,d^{2}-1$, we denote the generators of
${\mathrm{SU}}(d)$ which satisfy $\tr(\blm_{n})=0$ and
\begin{equation}
\tr\bigl(\blm_{m}\blm_{n}\bigr)=2{\,}\delta_{mn}
\ . \label{lij2}
\end{equation}
The factor $2$ in (\ref{lij2}) is rather traditional. Arbitrary
density operator can be represented in the form
\cite{bengtsson,kgka05}
\begin{equation}
\bro=\frac{1}{d}\left(
\pen+
\sum\nolimits_{n=1}^{d^{2}-1}
r_{n}\blm_{n}
\right)
{\>}, \label{brx}
\end{equation}
where $r_{n}=(d/2){\,}\tr\bigl(\bro\blm_{n}\bigr)$. These
components form a $(d^{2}-1)$-dimensional real vector, which
represents the density matrix $\bro$. Although the definition of
the Bloch vector is simple, the space of the Bloch vectors for
$d$-level system is difficult to determine. Some general
properties of the Bloch-vector space are studied in
\cite{kimura03,bk03,zs03}. By calculations, we obtain
\begin{equation}
\tr(\bro^{2})=\frac{1}{d}+\frac{2}{d^{2}}{\>}\|r\|_{2}^{2}
\ , \label{trb2}
\end{equation}
where $\|r\|_{2}$ denotes the vector $2$-norm. The formula
(\ref{icpr}) is then represented as
\begin{equation}
C(\nc|\bro)=\frac{1}{d^{2}}+
\frac{2(ad^{3}-1)}{d^{3}(d^{2}-1)}{\>}\|r\|_{2}^{2}
\ . \label{icprb}
\end{equation}
For $d=2$, this result is reduced to (\ref{inc2}). For the
completely mixed state $\bro_{*}=\pen/d$, components of the
generalized Bloch vectors are all zero. Hence, the formula
(\ref{icprb}) directly leads to (\ref{ccms}). Thus, we have useful
expressions in terms of the generalized Bloch vector.

\section{Tsallis' and R\'{e}nyi's formulations for a single SIC-POVM}\label{sec4}

In this section, we obtain uncertainty relations for a single
general SIC-POVM in terms of its Tsallis and R\'{e}nyi entropies.
Entropic functions are convenient tools to measure an uncertainty
in quantum measurements \cite{ww10,brud11}. The R\'{e}nyi and
Tsallis entropies are especially important generalizations of the
Shannon entropy. For $\alpha>0\neq1$, the Tsallis $\alpha$-entropy
of probability distribution $\{p_{j}\}$ is defined by
\cite{tsallis}
\begin{equation}
H_{\alpha}(p):=\frac{1}{1-\alpha}{\,}\left(\sum\nolimits_{j} p_{j}^{\alpha}
- 1 \right)
{\>}. \label{tsaent}
\end{equation}
The special case $\alpha=2$ gives the so-called linear entropy
$H_{2}(p)=1-C(p)$. The right-hand side of (\ref{tsaent}) is
usually rewritten in terms of the $\alpha$-logarithm
\begin{equation}
\ln_{\alpha}(x):=\frac{x^{1-{\alpha}}-1}{1-{\alpha}}
\ , \label{aldf}
\end{equation}
where $\alpha>0\neq1$ and $x>0$. The Tsallis $\alpha$-entropy
reads
\begin{equation}
H_{\alpha}(p)=-\sum\nolimits_{j}p_{j}^{\alpha}{\,}\ln_{\alpha}(p_{j})=
\sum\nolimits_{j}p_{j}{\>}{\ln_{\alpha}}{\left(\frac{1}{p_{j}}\right)}
{\,}. \label{tsaent2}
\end{equation}
In statistical physics, the entropy (\ref{tsaent}) was originally
introduced in \cite{tsallis}. Taking $\alpha\to1$, the
$\alpha$-logarithm is reduced to the standard logarithm.
Then the formula (\ref{tsaent}) gives the Shannon entropy
$H_{1}(p)=-\sum_{j}p_{j}\ln{p}_{j}$. Functional properties of the
entropy (\ref{tsaent}) and its conditional versions are considered
in \cite{sf06,rastkyb}.

For the given SIC-POVM $\nc=\{\nm_{j}\}$, the entropy
$H_{\alpha}(\nc|\bro)$ is obtained by substituting the
probabilities (\ref{jopr}) into the formula (\ref{tsaent}). It
turns out that these entropies are bounded from below. We will
derive lower bounds on the Tsallis $\alpha$-entropy for
$\alpha\in(0;2]$.

\newtheorem{prpn2}[prpn1]{Proposition}
\begin{prpn2}\label{prp2}
Let general SIC-POVM $\nc$ be characterized by the parameter $a$
in the sense of (\ref{ficnd}). For $\alpha\in(0;2]$ and arbitrary
density matrix $\bro$, the Tsallis $\alpha$-entropy satisfies the
state-dependent bound
\begin{equation}
H_{\alpha}(\nc|\bro)\geq
{\ln_{\alpha}}{\left(
\frac{d(d^{2}-1)}{(ad^{3}-1){\,}\tr(\bro^{2})+d(1-ad)}
\right)}
{\,}. \label{gsp1t}
\end{equation}
\end{prpn2}

{\bf Proof.} The following point was noticed in \cite{rast13b}. For
$\alpha\in(0;2]$ and arbitrary probability distribution, the
Tsallis $\alpha$-entropy obeys
\begin{equation}
H_{\alpha}(p)\geq{\ln_{\alpha}}{\left(\frac{1}{C(p)}\right)}
{\,}. \label{conc1}
\end{equation}
This formula is a direct consequence of Jensen's inequality for
the function $x\mapsto\ln_{\alpha}(1/x)$. Indeed, this function is
convex for $\alpha\in(0;2]$. Combining  (\ref{icpr}) with
(\ref{conc1}) immediately gives the claim (\ref{gsp1t}).
$\square$

For all $\alpha\in(0;2]$, the result (\ref{gsp1t}) provides a
state-dependent lower bound on the Tsallis $\alpha$-entropy of
probability distribution generated by a general SIC-POVM. For
$\alpha=2$, the inequality (\ref{gsp1t}) is always saturated.
Using (\ref{icprb}), we rewrite the bound (\ref{gsp1t}) in terms
of the generalized Bloch vector, namely
\begin{equation}
H_{\alpha}(\nc|\bro)\geq
{\ln_{\alpha}}{\left(
\frac{d^{3}(d^{2}-1)}{2(ad^{3}-1)\|r\|_{2}^{2}+d(d^{2}-1)}
\right)}
{\,}, \label{gsp1tb}
\end{equation}
where $\alpha\in(0;2]$. For $\alpha=1$, we obtain the lower bound
on the Shannon entropy, namely
\begin{equation}
H_{1}(\nc|\bro)\geq
{\ln}{\left(
\frac{d(d^{2}-1)}{(ad^{3}-1){\,}\tr(\bro^{2})+d(1-ad)}
\right)}
{\,}. \label{gsp1s}
\end{equation}
With a pure state $\bro=|\psi\rangle\langle\psi|$, the entropic
bound (\ref{gsp1t}) is reduced to the inequality
\begin{equation}
H_{\alpha}(\nc|\psi)\geq{\ln_{\alpha}}{\left(\frac{d(d+1)}{ad^{2}+1}\right)}
{\,}. \label{gsp1tp}
\end{equation}
For impure states, we have a stronger lower bound (\ref{gsp1t}).
The latter follows from increasing of the $\alpha$-logarithm and
and the fact that $\tr\bigl(\bro^{2}\bigr)<1$ for an impure state.
Here, we see a natural dependence on the measured state. The
right-hand side of (\ref{gsp1t}) reaches its maximum with the
completely mixed state $\bro_{*}=\pen/d$. Using
$\tr(\bro_{*}^{2})=d^{-1}$, the formula (\ref{gsp1t}) becomes
\begin{equation}
H_{\alpha}(\nc|\bro_{*})\geq
\ln_{\alpha}\bigl(d^{2}\bigr)
{\>}. \label{gsp1tm}
\end{equation}
The bound (\ref{gsp1tm}) is just saturated. Indeed, for all
$j=1,\ldots,d^{2}$ we have $\tr(\nm_{j})=d^{-1}$ and, herewith,
$p_{j}(\nc|\bro_{*})=d^{-2}$. Substituting this probability into
the right-hand side of (\ref{tsaent2}), we actually obtain
(\ref{gsp1tm}) with the sign of equality. With the completely
mixed state, the equality takes place for all $\alpha>0$
irrespectively to the parameter $a$. In the mentioned sense, the
state-dependent bound (\ref{gsp1t}) is tight. At the same time, we
proved (\ref{gsp1t}) only for $\alpha\in(0;2]$. Formulating lower
bounds on the entropy $H_{\alpha}(\nc|\bro)$ for $\alpha>2$ is an
open question.

To consider detection inefficiencies, we will use the following
approach \cite{rast13b}. To the given value $\eta\in[0;1]$ and
probability distribution $\{p_{j}\}$, one assigns a ``distorted''
distribution:
\begin{equation}
p_{j}^{(\eta)}=\eta{\,}p_{j}
\ , \qquad
p_{\varnothing}^{(\eta)}=1-\eta
\ . \label{dspd}
\end{equation}
The probability $p_{\varnothing}^{(\eta)}$ corresponds to the
no-click event. The parameter $\eta\in[0;1]$ characterizes a
detector efficiency. As was shown in the paper \cite{rastqqt}, for
all $\alpha>0$ we have
\begin{equation}
{H_{\alpha}}{\bigl(p^{(\eta)}\bigr)}=\eta^{\alpha}H_{\alpha}(p)+h_{\alpha}(\eta)
\ . \label{qtlm0}
\end{equation}
Here, the binary Tsallis entropy $h_{\alpha}(\eta)$ is expressed by
\begin{equation}
h_{\alpha}(\eta):=-\eta^{\alpha}\ln_{\alpha}(\eta)-(1-\eta)^{\alpha}\ln_{\alpha}(1-\eta)
\ . \label{bnta}
\end{equation}
Such results have been used in studying entropic Bell inequalities
with detector inefficiencies \cite{rchtf12}. Entropic uncertainty
relations with detection inefficiencies for mutually unbiased
bases were derived in \cite{rast13b}. Combining (\ref{gsp1t}) with
(\ref{qtlm0}), for $\alpha\in(0;2]$ we have
\begin{align}
H_{\alpha}^{(\eta)}(\nc|\bro)
&\geq\eta^{\alpha}{\,}
{\ln_{\alpha}}{\left(
\frac{d(d^{2}-1)}{(ad^{3}-1){\,}\tr(\bro^{2})+d(1-ad)}
\right)}
\nonumber\\
&+h_{\alpha}(\eta)
{\>}. \label{gsp1tt}
\end{align}
The entropy $H_{\alpha}^{(\eta)}(\nc|\bro)$ is calculated for the
distribution (\ref{dspd}), in which the initial distribution is
generated according to (\ref{jopr}). The result (\ref{gsp1tt}) is
an entropic uncertainty relation for a general SIC-POVM in the
model of detection inefficiencies. The inefficiency-free lower
bound (\ref{gsp1t}) is multiplied by factor $\eta^{\alpha}$ and
also added by the binary entropy $h_{\alpha}(\eta)$. Thus, an
additional uncertainty is caused by the detector.

Let us consider entropic bounds for a single general
SIC-POVM in terms of R\'{e}nyi's entropy. For $\alpha>0\neq1$,
the R\'{e}nyi $\alpha$-entropy of probability distribution
$\{p_{j}\}$ is defined as \cite{renyi61}
\begin{equation}
R_{\alpha}(p):=
\frac{1}{1-\alpha}{\ }{\ln}{\left(\sum\nolimits_{j} p_{j}^{\alpha}
\right)}
{\>}. \label{renent}
\end{equation}
When $\alpha\to1$, this expression is reduced to the
standard Shannon entropy. The entropy (\ref{renent}) is a
non-increasing function of order $\alpha$ \cite{renyi61}.
Taking $\alpha=2$, the expression (\ref{renent}) gives the
collision entropy
\begin{equation}
R_{2}(p)={-\ln}{\left(\sum\nolimits_{j}p_{j}^{2}\right)}
=-\ln{C}(p)
{\>}. \label{clen}
\end{equation}
Note that the collision entropy is closely related to the index of
coincidence and the linear entropy. In the limit
$\alpha\to\infty$, we have the so-called min-entropy
\begin{equation}
R_{\infty}(p)=-\ln\bigl(\max{p}_{j}\bigr)
\ . \label{mnen}
\end{equation}
The min-entropy is of particular interest in cryptography
\cite{ngbw12}. It is also related to the extrema of the
discrete Wigner function \cite{MWB10}. R\'{e}nyi-entropies
uncertainty relations are significant in studying the connection
between complementarity and uncertainty principles
\cite{bosyk13a}. Using the R\'{e}nyi entropy, the writers of
\cite{rprz12} formulated trade-off relations for a
trace-preserving quantum operation. An extension of such trade-off
relations in terms of the so-called unified entropies was
discussed in \cite{rast13a}. For a SIC-POVM $\nc=\{\nm_{j}\}$, the
entropy $R_{\alpha}(\nc|\bro)$ is obtained by substituting the
probabilities (\ref{jopr}) into (\ref{renent}). We now consider
lower bounds on this entropy.

\newtheorem{prpn3}[prpn1]{Proposition}
\begin{prpn3}\label{prp3}
Let general SIC-POVM $\nc$ be characterized by the parameter $a$
in the sense of (\ref{ficnd}). For $\alpha\in[2;\infty)$ and
arbitrary density matrix $\bro$, the R\'{e}nyi $\alpha$-entropy
satisfies the state-dependent bound
\begin{align}
&R_{\alpha}(\nc|\bro)\geq
\nonumber\\
&\frac{\alpha}{2(\alpha-1)}{\ }
{\ln}{\left(
\frac{d(d^{2}-1)}{(ad^{3}-1){\,}\tr(\bro^{2})+d(1-ad)}
\right)}
{\,}. \label{gsp1r}
\end{align}
\end{prpn3}

{\bf Proof.} For $\alpha\geq2$ and arbitrary probability
distribution, we write the inequality
\begin{equation}
\left(\sum\nolimits_{j}p_{j}^{\alpha}\right)^{1/\alpha}
\leq\left(\sum\nolimits_{j}p_{j}^{2}\right)^{1/2}=C(p)^{1/2}
\ . \label{nmq2}
\end{equation}
This inequality follows from theorem 19 of the book \cite{hardy}.
The function $x\mapsto(1-\alpha)^{-1}\ln{x}$ decreases for
$\alpha>1$. Combining this with (\ref{renent}) and (\ref{nmq2})
further gives
\begin{equation}
R_{\alpha}(p)\geq\frac{\alpha}{2(1-\alpha)}{\>}\ln{C}(p)
\ . \label{rangc}
\end{equation}
The formulas (\ref{icpr}) and (\ref{rangc}) completes the
proof. $\square$

The formula (\ref{gsp1r}) provides a state-dependent lower bound
on the R\'{e}nyi $\alpha$-entropy of probability distribution
generated by a general SIC-POVM. Due to (\ref{icprb}), we can
rewrite the bound (\ref{gsp1t}) in the form
\begin{equation}
R_{\alpha}(\nc|\bro)\geq\frac{\alpha}{2(1-\alpha)}{\ }
{\ln}{\left(
\frac{1}{d^{2}}+
\frac{2(ad^{3}-1)}{d^{3}(d^{2}-1)}{\>}\|r\|_{2}^{2}
\right)}
{\,}. \label{gsp1rb}
\end{equation}
For $\alpha=2$, the inequality (\ref{gsp1r}) gives a bound on the
collision entropy written as
\begin{equation}
R_{2}(\nc|\bro)\geq{\ln}{\left(
\frac{d(d^{2}-1)}{(ad^{3}-1){\,}\tr(\bro^{2})+d(1-ad)}
\right)}
{\,}. \label{asrlb2}
\end{equation}
As the R\'{e}nyi $\alpha$-entropy does not increase with $\alpha$,
the bound (\ref{asrlb2}) is valid for all R\'{e}nyi's entropies of
order $\alpha\in(0;2]$, including the Shannon-entropy case
(\ref{gsp1s}). For $\bro=|\psi\rangle\langle\psi|$, the entropic
bound (\ref{gsp1r}) is reduced to its pure-state form
\begin{equation}
R_{\alpha}(\nc|\psi)\geq
\frac{\alpha}{2(\alpha-1)}{\ }
{\ln}{\left(\frac{d(d+1)}{ad^{2}+1}\right)}
{\,}, \label{gsp1rp}
\end{equation}
where $\alpha\in[2;\infty)$. Due to $\tr\bigl(\bro^{2}\bigr)<1$
and increase of the logarithm, the lower bound (\ref{gsp1rp}) is
weaker than (\ref{gsp1r}). The right-hand side of (\ref{gsp1r}) is
maximal for $\bro_{*}=\pen/d$. It is easy to see that
$p_{j}(\nc|\bro_{*})=d^{-2}$ and, therefore,
$R_{\alpha}(\nc|\bro_{*})=2\ln{d}$ for all $\alpha>0$. With the
completely mixed state, the right-hand side of (\ref{asrlb2})
actually gives the bound $2\ln{d}$ for $\alpha\in(0;2]$. In this
sense, the derived bound (\ref{gsp1r}) is tight for such values of
$\alpha$. For $\alpha>2$, the result (\ref{gsp1r}) is always
approximate. In fact, it gives the inequality
$R_{\infty}(\nc|\bro_{*})\geq\ln{d}$, which contains only a half
of the exact value $R_{\infty}(\nc|\bro_{*})=2\ln{d}$. A way to
improve relations with the min-entropy was discussed in
\cite{rast13b}. We now extend such a treatment to general
SIC-POVMs.

\newtheorem{prpn4}[prpn1]{Proposition}
\begin{prpn4}\label{prp4}
Let general SIC-POVM $\nc$ be characterized by the parameter $a$
in the sense of (\ref{ficnd}). For arbitrary density matrix
$\bro$, the min-entropy satisfies the state-dependent bound
\begin{align}
&R_{\infty}(\nc|\bro)\geq
\nonumber\\
&2\ln{d}-{\ln}{\left(1+\sqrt{ad^{3}-1}\sqrt{\tr(\bro^{2}){\,}d-1}\right)}
{\,}. \label{mest}
\end{align}
\end{prpn4}

{\bf Proof.} In appendix A of the paper \cite{rast13b}, we proved
the following statement. If the $n$ positive numbers $x_{j}$
obey the conditions $\sum_{j=1}^{n}x_{j}=1$ and
$\sum_{j=1}^{n}x_{j}^{2}=b^{2}$, then
\begin{equation}
\underset{1\leq{j}\leq{n}}{\max}
{\,}x_{j}\leq\frac{1}{n}\left(1+\sqrt{n-1}\sqrt{n{b}^{2}-1}\right)
{\,}. \label{aust}
\end{equation}
Substituting (\ref{icpr}) instead of $b^{2}$ and $d^{2}$ instead
of $n$ into (\ref{aust}), we finally obtain
\begin{equation}
\underset{1\leq{j}\leq{d}^{2}}{\max}
{\,}p_{j}(\nc|\bro)\leq
\frac{1}{d^{2}}\left(1+\sqrt{ad^{3}-1}\sqrt{\tr(\bro^{2}){\,}d-1}\right)
{\,}. \label{mest1}
\end{equation}
Combining (\ref{mnen}) with (\ref{mest1}) directly leads to
(\ref{mest}), since the function $x\mapsto-\ln{x}$ decreases.
$\square$

The lower bound (\ref{mest}) is clearly stronger than the limiting
case $\alpha\to\infty$ of the right-hand side of (\ref{gsp1r}).
With the completely mixed state, the result (\ref{mest}) gives the
tight bound $2\ln{d}$ due to $\tr(\bro_{*}^{2})=d^{-1}$. The
right-hand side of (\ref{mest}) increases as the quantity
$\tr(\bro^{2})$ decreases. In other words, the more a state is
mixed, the more the bound (\ref{mest}). Replacing $\tr(\bro^{2})$
with $1$, the right-hand side of (\ref{mest}) gives the lower
bound for the pure-state case. For $a=d^{-2}$, the upper bound
(\ref{mest1}) on the maximal probability leads to the analogous
bound for a usual SIC-POVM. This particular case was already discussed
in \cite{rast13b}.

\section{Uncertainty relations of the Maassen--Uffink type}\label{sec6}

In this section, we will discuss entropic uncertainty relations
for a pair of general SIC-POVMs. Since the celebrated Heisenberg's
result was published \cite{heisenberg}, many approaches to
incompatibilities in quantum measurements have been proposed.
Entropic uncertainty relations were studied in many important
cases \cite{ww10,brud11}. Results of such a kind are mainly based
on the Maasen--Uffink approach \cite{maass}. This approach has
been developed with use of various entropic functions. Entropic
bounds in terms of generalized entropies entropic bounds were
utilized in studying many topics such as the case of conjugate
observables \cite{birula06,pp96}, quantifying number-phase
uncertainties \cite{rast105,rast12num}, incompatibilities of
anti-commuting observables \cite{ww08} and reformulations in
quasi-Hermitian models \cite{rast12quasi}. In the context of
simultaneous measurements of complementary observables,
uncertainty relations in terms of both the Tsallis and R\'{e}nyi
entropies are examined in \cite{luis13}. The method of Maassen and Uffink
uses the Riesz theorem. It therefore leads to lower bound on the sum of two
entropies, whose orders obey a certain condition
\cite{birula06,rast12num}. Entropic inequalities for
quantum tomograms of qudit states are examined in \cite{mman13}.
Recently, new universal approach to entropic uncertainty relations
has been proposed \cite{fgg13,prz13}. Apparently, this approach
will play a significant role in future research. We will formulate
uncertainty relations for two general SIC-POVMs in terms of the
R\'{e}nyi and Tsallis entropies as well as their symmetrized
versions. We have the following statement.

\newtheorem{prpn5}[prpn1]{Proposition}
\begin{prpn5}\label{prp5}
Let $\mc=\{\mm_{i}\}$ and $\nc=\{\nm_{j}\}$ be general SIC-POVMs.
To any density matrix $\bro$, we assign the quantity
\begin{equation}
g(\mc,\nc|\bro):=\underset{1\leq{i},j\leq{d}^{2}}{\max}
{\,}\frac{\bigl|\tr(\mm_{i}\nm_{j}\bro)\bigr|}{p_{i}(\mc|\bro)^{1/2}{\,}p_{j}(\nc|\bro)^{1/2}}
{\ }. \label{gmndf}
\end{equation}
Let positive orders $\alpha$ and $\beta$ obey
$1/\alpha+1/\beta=2$, and let $\mu=\max\{\alpha,\beta\}$. Then the
corresponding Tsallis entropies satisfy the inequality
\begin{equation}
H_{\alpha}(\mc|\bro)+H_{\beta}(\nc|\bro)\geq{\ln_{\mu}}{\Bigl(g(\mc,\nc|\bro)^{-2}\Bigr)}
{\>}. \label{mnts1}
\end{equation}
Under the same preconditions, the corresponding R\'{e}nyi
entropies satisfy the inequality
\begin{equation}
R_{\alpha}(\mc|\bro)+R_{\beta}(\nc|\bro)\geq
-2\ln{g}(\mc,\nc|\bro)
\ . \label{mnrn1}
\end{equation}
\end{prpn5}

The presented formulations (\ref{mnts1}) and (\ref{mnrn1})
immediately follows from the results of section 3 of
\cite{rast104}. The quantity (\ref{gmndf}) explicitly depends on
the pre-measurement density matrix $\bro$. It is of certain
interest to obtain the state-independent form of entropic bounds
\cite{deutsch}. For the entropic relations (\ref{mnts1}) and
(\ref{mnrn1}), a way to obtain such forms was considered in
\cite{rast104,rast12quasi}. Using the Cauchy-Schwarz inequality
for the Hilbert--Schmidt inner product and the inequality
(\ref{schqq}), we finally obtain
\begin{align}
& g(\mc,\nc|\bro)\leq\bar{f}(\mc,\nc)
\ . \label{gbfmn0}\\
& \bar{f}(\mc,\nc):=\underset{1\leq{i},j\leq{d}^{2}}{\max}
{\,}\bigl\|\mm_{i}^{1/2}\bigr\|_{\infty}\bigl\|\nm_{j}^{1/2}\bigr\|_{\infty}
{\>}. \label{gbfmn}
\end{align}
Here, we used three formulas
\begin{align}
& p_{i}(\mc|\bro)=\bigl\|\mm_{i}^{1/2}\sqrt{\bro}\bigr\|_{2}^{2}
\ , \quad
p_{j}(\nc|\bro)=\bigl\|\nm_{j}^{1/2}\sqrt{\bro}\bigr\|_{2}^{2}
\ , \nonumber\\
& \tr(\mm_{i}\nm_{j}\bro)=\bigl\langle\mm_{i}\sqrt{\bro}{\,},\nm_{j}\sqrt{\bro}\bigr\rangle_{\mathrm{hs}}
\ . \label{3rel}
\end{align}
Due to (\ref{gbfmn}), we obtain the state-independent entropic
bounds
\begin{align}
H_{\alpha}(\mc|\bro)+H_{\beta}(\nc|\bro)&\geq{\ln_{\mu}}{\Bigl(\bar{f}(\mc,\nc)^{-2}\Bigr)}
{\>}, \label{mnts10}\\
R_{\alpha}(\mc|\bro)+R_{\beta}(\nc|\bro)&\geq
-2\ln\bar{f}(\mc,\nc)
{\>}, \label{mnrn10}
\end{align}
in which the parameters $\alpha$, $\beta$, and $\mu$ are defined
as in Proposition \ref{prp5}. We can also reformulate these
uncertainty relations in terms of the parameters $a_{\mc}$ and
$a_{\nc}$ defined according to (\ref{ficnd}). Using the inequality
(\ref{npqr}), we write
\begin{equation}
\|\mm_{i}\|_{\infty}{\,}\|\nm_{j}\|_{\infty}\leq\|\mm_{i}\|_{2}{\,}\|\nm_{j}\|_{2}=\sqrt{a_{\mc}a_{\nc}}
{\ }. \label{naman}
\end{equation}
This inequality is always saturated with the usual SIC-POVMs, when
$a_{\mc}a_{\nc}=d^{-4}$. By positivity, we also have
\begin{equation}
\bar{f}(\mc,\nc)^{2}=
\underset{1\leq{i},j\leq{d}^{2}}{\max}
{\,}\|\mm_{i}\|_{\infty}{\,}\|\nm_{j}\|_{\infty}
\ . \label{4rel}
\end{equation}
Combining (\ref{4rel}) with (\ref{naman}) finally gives
\begin{equation}
g(\mc,\nc|\bro)^{-2}\geq
\bar{f}(\mc,\nc)^{-2}\geq{a}_{\mc}^{-1/2}a_{\nc}^{-1/2}
{\ }. \label{naman1}
\end{equation}
Since the function $x\mapsto\ln_{\mu}(x)$ is increasing, the
inequality (\ref{naman1}) leads to entropic uncertainty relations
\begin{align}
H_{\alpha}(\mc|\bro)+H_{\beta}(\nc|\bro)&\geq{\ln_{\mu}}{\Bigl(a_{\mc}^{-1/2}a_{\nc}^{-1/2}\Bigr)}
{\>}, \label{mnts11}\\
R_{\alpha}(\mc|\bro)+R_{\beta}(\nc|\bro)&\geq
-\frac{1}{2}{\,}\bigl(\ln{a}_{\mc}+\ln{a}_{\nc}\bigr)
{\>}, \label{mnrn11}
\end{align}
which follow from (\ref{mnts10}) and (\ref{mnrn10}), respectively.
The parameters $a_{\mc}$ and $a_{\nc}$ range according to
(\ref{resa}). For two usual SIC-POVMs, the right-hand sides of
(\ref{mnts11}) and (\ref{mnrn11}) take values
$\ln_{\mu}\bigl(d^{2}\bigr)$ and $2\ln{d}$ due to
$a_{\mc}=a_{\nc}=d^{-2}$. For other general SIC-POVMs, these
bounds are strictly stronger. They may increase up
to $\ln_{\mu}\bigl(d^{3}\bigr)$ and $3\ln{d}$, when
$a_{\mc}$ and $a_{\nc}$ both reach $d^{-3}$. However, the latter 
does not lead to legitimate SIC-POVMs. Since a SIC-POVM has $d^{2}$
different outcomes, its Tsallis and Renyi $\alpha$-entropies are
bounded from above by $\ln_{\alpha}\bigl(d^{2}\bigr)$
and $2\ln{d}$, respectively. For two general SIC-POVMs, the formulas
(\ref{mnts11}) and (\ref{mnrn11}) show that the sum of the two
corresponding entropies is not less than the maximal possible
value for one of them. Thus, we have obtained non-trivial entropic
bound of the Maassen--Uffink type.

Finally, we discuss reformulations in terms of the 
symmetrized entropies. For a pair of observables, uncertainty
relations in terms of symmetrized entropies were given in both the
R\'{e}nyi \cite{birula06} and Tsallis formulations \cite{raja95}.
In \cite{rast13b}, lower bounds on the sum of symmetrized
entropies have been derived for mutually unbiased bases.
We assume that the entropic orders obey $1/\alpha+1/\beta=2$.
Using $s\in[0;1)$, we parameterize the orders as
\begin{equation}
\max\{\alpha,\beta\}=\frac{1}{1-s}
\ , \qquad
\min\{\alpha,\beta\}=\frac{1}{1+s}
\ .  \label{prms}
\end{equation}
The symmetrized Tsallis and R\'{e}nyi entropies are respectively
defined by
\begin{align}
\wrh_{s}(\nc|\bro)&:=\frac{1}{2}{\,}\Bigl(H_{\alpha}(\nc|\bro)+H_{\beta}(\nc|\bro)\Bigr)
{\>}, \label{hsm}\\
\wrn_{s}(\nc|\bro)&:=\frac{1}{2}{\,}\Bigl(R_{\alpha}(\nc|\bro)+R_{\beta}(\nc|\bro)\Bigr)
{\>}. \label{rsm}
\end{align}
Taking $\mu=(1-s)^{-1}$, for the sum
$\wrh_{s}(\mc|\bro)+\wrh_{s}(\nc|\bro)$ we have the three lower
bounds (\ref{mnts1}), (\ref{mnts10}), and (\ref{mnts11}).
Similarly, the lower bounds (\ref{mnrn1}), (\ref{mnrn10}), and
(\ref{mnrn11}) are all valid for the sum
$\wrn_{s}(\mc|\bro)+\wrn_{s}(\nc|\bro)$. Using symmetrized
entropies, we can extend bounds of the Maassen--Uffink type to
more than two measurements. An example with several mutually
unbiased bases has been analyzed in \cite{rast13b}. In principle,
this idea could be applied to general SIC-POVMs.

\section{Conclusions}\label{sec7}

We have reported some properties of general symmetric
informationally complete POVMs. SIC-POVMs are of interest in
various topics such as quantum state tomography and quantum
cryptography. Thus, the presented results may be useful within
quantum technologies. For a general SIC-POVM and arbitrary
measured state, the index of coincidence of generated probability
distribution is exactly calculated. This result is a
generalization of the previous calculation for a rank-one
SIC-POVM. The obtained index of coincidence is expressed in terms
of dimensionality, the trace of squared density matrix, and a 
parameter characterizing the given SIC-POVM. The trace of
squared density matrix is one of measures quantifying a degree of
state impurity. The calculation of the index of coincidence leads
to entropic uncertainty relations for a single general SIC-POVM.
We have expressed state-dependent formulations in terms of both
the R\'{e}nyi and Tsallis entropies. These formulations are an
extension of entropic relations previously given in
\cite{rast13b}. The min-entropy uncertainty relation is separately
considered. For a pair of general SIC-POVMs, we discussed
uncertainty relations of the Maassen-Uffink type. Reformulations
in terms of the symmetrized entropies are briefly considered. A
new important approach to obtaining entropic uncertainty bounds
with the use of majorization technique has recently been proposed
in the papers \cite{fgg13,prz13}. It may be interesting to study
uncertainty relations for general SIC-POVMs on the base of 
majorization techniques.

\medskip

The author is grateful to Karol \.{Z}yczkowski for useful
correspondence and valuable comments. The author thanks anonymous
referee for constructive criticism.

\end{document}